# Variations of petrophysical properties and spectral induced polarization in response to drainage and imbibition: a study on a correlated random tube network

Alexis Maineult [1,*], Damien Jougnot [1], and André Revil [2,3]

1) Sorbonne Universités, UPMC Univ Paris 06, CNRS, EPHE, UMR 7619 Metis, 4 place Jussieu, 75005 Paris, France
2) Univ. Grenoble Alpes, CNRS, IRD, IFSTTAR, ISTerre, 38000 Grenoble, France
3) Univ. Savoie Mont Blanc, ISTerre, 73000 Chambery, France

**Short title:** drainage and imbibition on tube network

* Corresponding author. E-mail: alexis.maineult@upmc.fr, tel: +33 (0)1 44 27 43 36

**SUMMARY**

We implement a procedure to simulate the drainage and imbibition in random, two-dimensional, square networks. We compute the resistivity index, the relative permeability, and the characteristic lengths of a correlated network at various saturation states, under the assumption that the surface conductivity can be neglected. These parameters exhibit a hysteretic behaviour. Then, we calculate the Spectral Induced Polarization (SIP) response of the medium, under the assumption that the electrical impedance of each tube follows a local Warburg conductivity model, with identical DC conductivity and chargeability for all the tubes. We evidence that the shape of the SIP spectra depends on the saturation state. The analysis of the evolution of the macroscopic Cole-Cole parameters of the spectra in function of the saturation also behaves hysteretically, except for the Cole-Cole exponent. We also observe a power-law relationship between the macroscopic DC conductivity and time constant and the relative permeability. We also show that the frequency peak of the phase spectra is directly related to the characteristic length and to the relative permeability, underlining the potential interest of SIP measurements for the estimation of the permeability of unsaturated media.

## 1 INTRODUCTION

Geophysical methods are now commonly used to characterize the hydrological state of the subsurface, for instance to localize aquifers (e.g., Vereecken *et al.* 2006; Hubbard & Linde 2011; Binley *et al.* 2015) and to determine their hydraulic conductivity and storage coefficient distributions (e.g., Soueid Ahmed *et al.* 2016a, b). Electrical resistivity tomography (ERT) provides some information regarding the architecture of the subsurface (e.g., Binley & Kemna 2005; Johnson et al. 2010) and this method is sensitive to the water content, temperature, salinity, and cation exchange capacity (Shainberg *et al.* 1980; Sen *et al.* 1981). To date, one remaining challenge for field hydrogeophysicists is to be able to estimate, in a reliable way, the saturation state of the medium and its permeability at partial saturation, which are parameters of importance for the characterization of the vadose zone (Doussan & Ruy, 2009).

In the past decade, the induced polarization (IP) method has shown its potential to estimate the permeability in saturated conditions (Slater and Lesmes 2002; Tong *et al.* 2004; Binley *et al.* 2005; Hördt *et al.* 2007; Weller *et al.* 2010; Attwa & Günther 2013; Slater *et al.* 2014; Revil *et al.* 2015). Laboratory studies have also evidenced that some characteristics of the IP response are directly modified when the saturation is changed (e.g. Titov *et al.* 2004; Cosenza *et al.* 2007; Jougnot *et al.* 2010; Breede *et al.* 2011, Schmutz *et al.* 2012).

The spectral induced polarization (SIP) method consists in injecting a sinusoidal electrical current $I = I_0 \sin(\omega t)$ ($\omega$ is the pulsation frequency and $t$ the time) in the medium with two injection electrodes, and simultaneously measuring the resulting electrical potential $\Delta V$ between two measurement electrodes. The complex conductivity spectrum is obtained by repeating the measurement for different frequencies and is given by:

$$\sigma^*(\omega) = \frac{1}{G}\frac{I^*(\omega)}{\Delta V^*(\omega)} = \frac{1}{G}\frac{|I^*(\omega)|}{|\Delta V^*(\omega)|}e^{-i\varphi(\omega)} = \sigma(\omega)e^{-i\varphi(\omega)} \qquad (1)$$

where the superscript * denotes a complex quantity. Here $\sigma^*$ stands for complex conductivity,

$\sigma$ for the conductivity amplitude, $\varphi$ for the phase, $G$ for the geometrical factor determined from the electrode locations and the boundary conditions, and $i$ for the pure imaginary number. The method of random tube networks has long been applied to study the permeability and the formation factor (e.g. Kirkpatrick 1973; Koplik 1981; David et al. 1990; David 1993, Bernabé 1995). Recently, Maineult *et al*. (2017) used the method to study the upscaling of the SIP response, from pore scale to sample scale and to analyse the relationship between the pore size distribution and the dispersion shown by the complex conductivity spectra. Briefly, a sinusoidal current is applied between the two end-faces of a 2D square network (Figure 1), made of a set of connected capillaries with identical length but different radius (Figure 2a, b). The conservation law for the complex electrical current (Kirchhoff, 1845) can be written at each node. Knowing the complex impedance of each tube, this produces a linear system whose resolution gives the complex electrical potential at each node. From the electrical potential and the tube impedances, the macroscopic current entering in or exiting from the system (note that they are equal) can be deduced. The complex conductivity is then computed using Eq. 1 ($G$ is equal to 1 in this case). To summarize, this methodology allows the macroscopic impedance of a set of local impedances to be calculated (Figure 2c). In this paper, we apply the same methodology to study the effect of a desaturation process (i.e., drainage) and then a resaturation process (i.e., imbibition) on the macroscopic SIP response of the network.

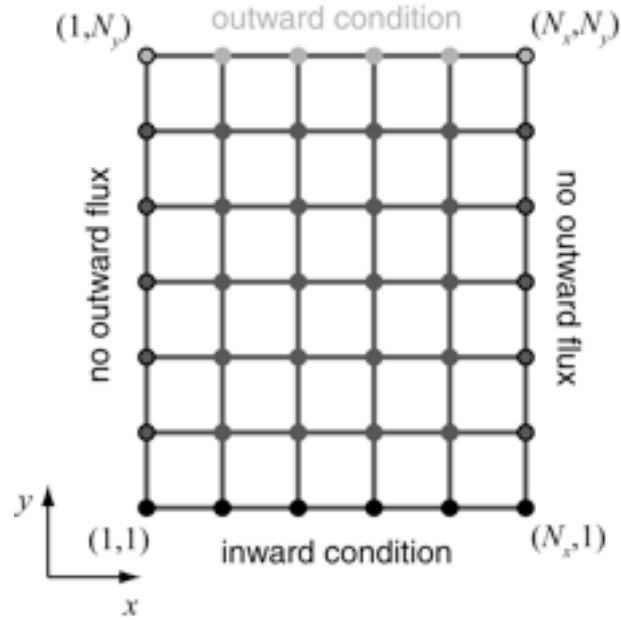

**Figure 1.** Scheme of a 2D square tube network for the computation of the macroscopic complex conductivity (CC), apparent formation factor (AFF) and apparent hydraulic permeability (AHP). There are no outflowing fluxes on the lateral faces (nodes $x = 1$ and $x = N_x$). The inward condition (nodes $y = 1$) is a potential equal to $V_0 e^{i\omega t}$ for CC, and 1 for AFF and AHP. The outward condition (nodes $y = N_y$) is a potential equal to 0 in the three cases.

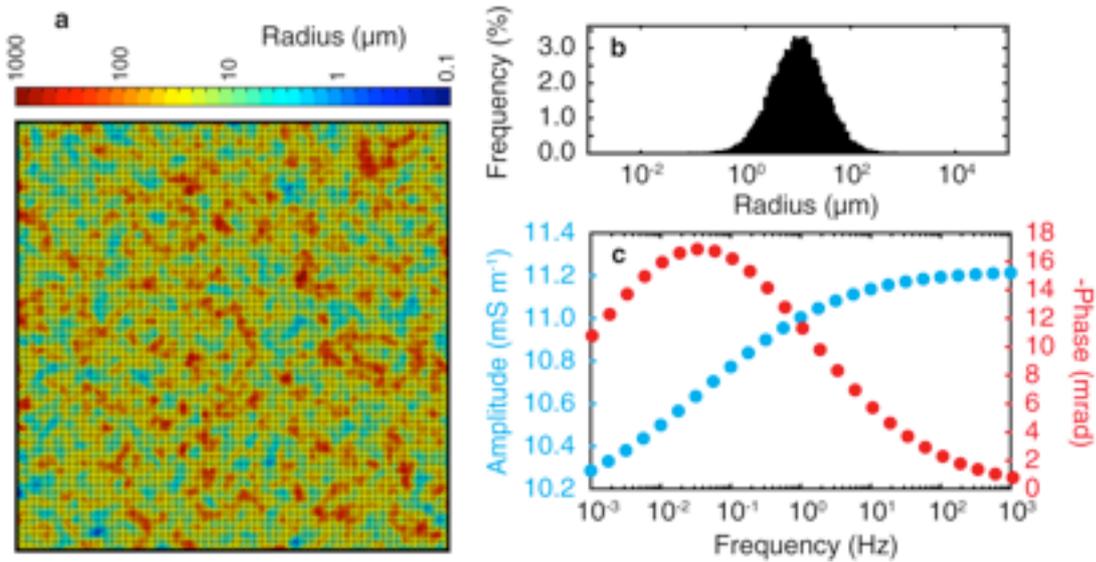

**Figure 2. a**: 100x100 studied correlated network used in this study (the decimal logarithm of the radii is normally distributed, with a mean radius of 10 μm and a standard deviation of 0.4942, a Hurst exponent of 0.9, and a characteristic scale $L_c$ of 1). **b**: associated radius distribution, and **c**: macroscopic complex conductivity response of the network (for a Warburg model with a DC conductivity of 10 mS m$^{-1}$, a chargeability of 0.1 V V$^{-1}$, and a diffusion coefficient of 10$^{-11}$ m$^2$ s$^{-1}$). See Maineult *et al.* (2017) for more details.

**2 METHOD**

We consider the 100×100 (i.e., 19800 tubes) network of Figure 2a, for which the logarithm of the pore radius distribution $r$ is normally-distributed, with a mean radius of 10 µm and a standard deviation of 0.4942. All tubes have the same length $l$. The medium is correlated, with a Hurst exponent (related to the fractal dimension of the medium, e.g., Turcotte 1997) of 0.9, and an isotropic characteristic length scale of 1. It corresponds to the correlated medium used in Maineult *et al.* (2017) (their Figure 5). In the present work, we consider the effect of partial saturation on the up-scaled SIP response. By applying the Young-Laplace equation, it is possible to relate an equivalent radius $r_{eq}$ (in m) that drains at a specific matric potential by (e.g., Jurin, 1717)

$$h = \frac{2\gamma \cos\theta}{\rho_w g r_{eq}}, \qquad (2)$$

where $h$ is the matric potential (in m), $\gamma$ is the surface tension of water (0.0727 N m$^{-1}$ at 20°C), $\theta$ is the contact angle (often considered to be 0°, which yields $\cos\theta = 1$, see Bear, 1972), $\rho_w$ is the water density (in kg m$^{-3}$), and $g$ is the gravitational acceleration (in m s$^{-2}$). This approach is traditionally used in soil science, when using simplified geometries to study the effect of water saturation upon effective properties: for hydrodynamic properties (e.g., Jury *et al.* 1991; Or & Tuller 1999; Guarracino 2007), hysteresis (e.g., Soldi *et al.* 2017), electrical conductivity (e.g., Niu *et al.* 2015) or self-potential (e.g., Packard 1953; Jougnot *et al.* 2012). However, given that the geometry of the porous medium is here known, we applied specific desaturation and re-saturation processes that are described below and in Figure 3. The hysteretic nature of the soil-water retention curve yields to different water content at a given matric potential during desaturation or resaturation of porous media (e.g. Mualem 1973). This hysteresis can be explained by different physical effects such as pore throats ("ink-bottle" effect), entrapped air, and possible changes in the contact angle of advancing or receding meniscus (see Pham *et al.* 2005 for a review). In this study, we consider the so-called "ink-bottle" effect: that is, parts of the network are not accessible to air or water entry due the pore size distribution as a result of Eq. (2).

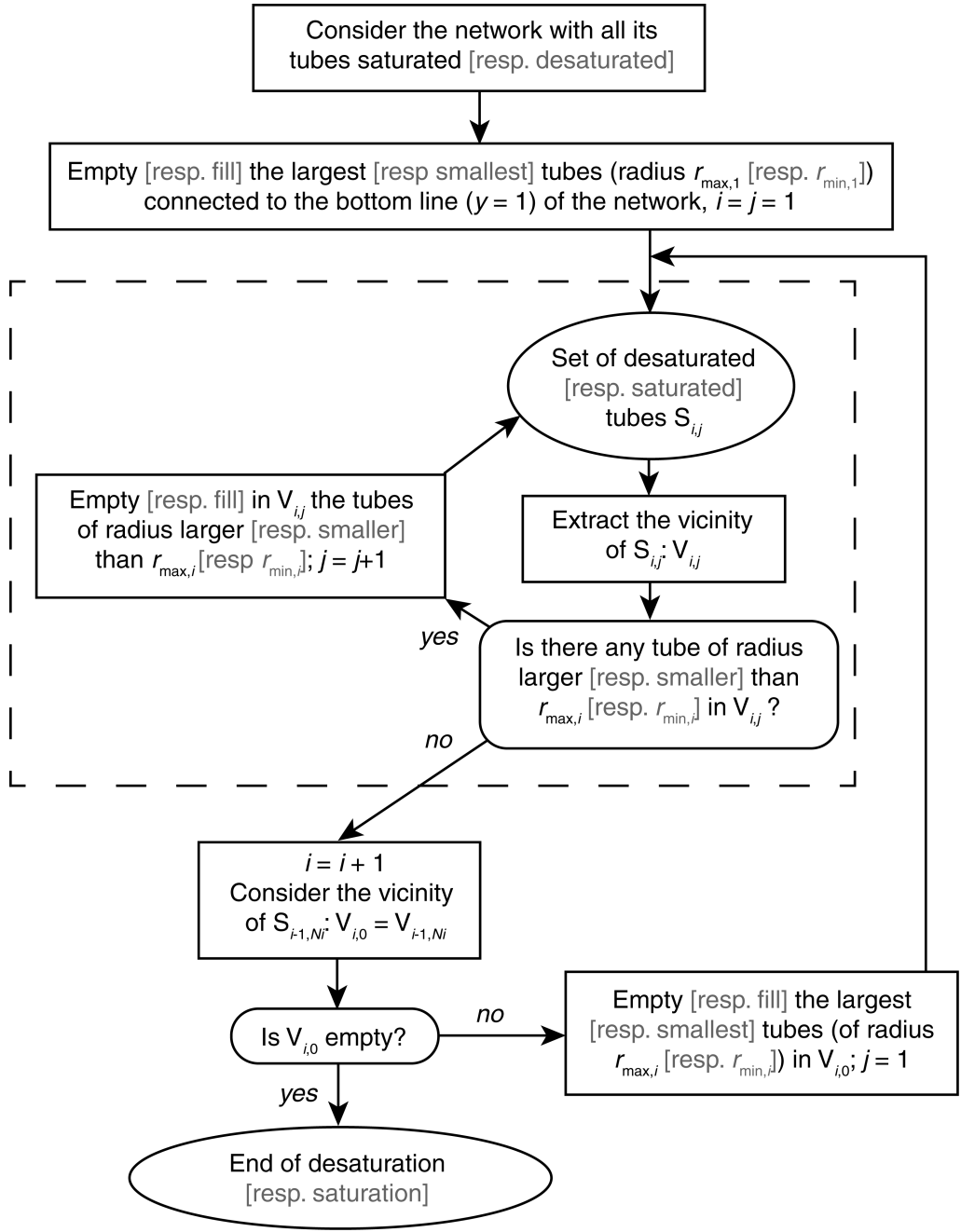

**Figure 3.** Diagram explaining the saturation and desaturation (grey words) of the network.

## 2.1 Desaturation of the tube network

Concerning drainage (desaturation), we consider the network initially fully saturated with water (Figure 4a, saturation $S_w$ of 100 %). We start the desaturation process from the bottom of the network (nodes $y = 1$, Figure 1), by emptying the largest tubes (or the largest tube, if there is only one) connected to this face. We denote their radius $r_{max,1}$, and the set of the so emptied tubes $S_{1,1}$ (step 0). We then consider the vicinity $V_{1,1}$ of $S_{1,1}$, i.e. the set of tubes connected to $S_{1,1}$; we empty the tubes with a radius larger than or equal to $r_{max,1}$ in $V_{1,1}$, to obtain $S_{1,2}$ (step 1). We continue the same process until the number of emptied tubes does not increase anymore. We note the set of emptied tubes after stabilization $S_{1,N1}$ (last step). To empty a tube, we just set its radius to 0, so the water saturation associated to the radius $r_{max,1}$ is simply computed at the end of the last step as:

$$S_w(r_{max,1}) = \frac{\sum_{i=1}^{N_t} r_i^2}{\Sigma_0} \qquad (3)$$

where $\Sigma_0$ is the sum of the squared radii for the complete network (i.e., at 100 % saturation), and $N_t$ the total number of tubes. Afterwards, we consider the vicinity of $S_{1,N1}$, i.e., $V_{1,N1} = V_{2,0}$. We empty in $V_{2,0}$ the largest tubes, whose diameter, denoted $r_{max,2}$, is obviously smaller than $r_{max,1}$. We repeat the procedure from step 1 to the last step to obtain $S_{2,N2}$. Finally, the entire process is repeated until the network is totally emptied.

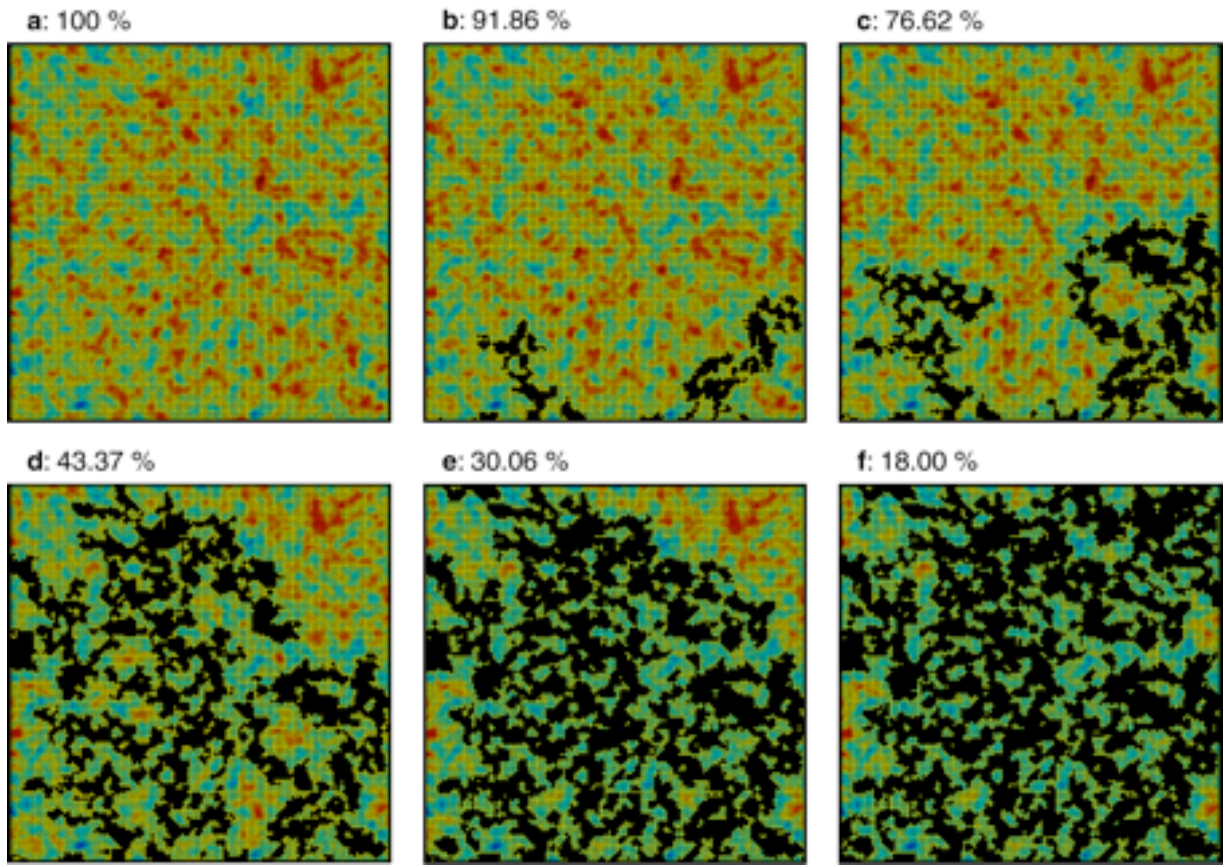

**Figure 4.** Example of saturation states during drainage. The largest radii are desaturated first (i.e., they are set to 0), starting from the bottom of the network (i.e., nodes $y = 1$, see Figure 1). The colour scale is the same as in Figure 2a.

Different desaturation states are shown in Figure 4. The desaturated zone forms clusters, which can appear very suddenly, when zones with radii largest than $r_{max,M}$ are connected to emptied tubes with radius equal to $r_{max,M}$. This phenomenon is known as "Haines jumps" (e.g., Haines 1930, DiCarlo *et al.* 2003, Haas & Revil 2009, Berg *et al.* 2013): when the air passes from a pore throat to a wider pores and displaces water, it generate a sudden drop in capillary pressure (see also Figure 6, on which the drops are clearly visible). Also, there is a critical saturation for which two opposite end-faces are disconnected (i.e., there is no continuous path for the wetting phase between them) – in this case, 30.06 % in the transversal direction (between faces $x = 1$ and $x = 100$, see Figure 1), and 18 % in the longitudinal direction (between faces $y = 1$ and $y = 100$). This difference is due to the fact that the desaturation process occurs along the longitudinal direction.

Finally, note that we continue the drainage process until $S_w = 0$, that is to say we do not consider water trapping in the tubes (their local saturation is 1 or 0), or in disconnected clusters of tubes. The desaturation process that we consider here is therefore similar to what we could obtain by putting the end-face $y = 1$ under vacuum.

**2.2 Saturation of the tube network**

For the imbibition (saturation), we start from a network with all its radii set to 0. Again, we start the saturation process from the bottom of the network (nodes $y = 1$, Figure 1). We open the smallest tubes (or the smallest tube, if there is only one) connected to this face – We denote their radius $r_{min,1}$, and the set of opened tubes is expressed as $S_{1,1}$ (step 0). Considering the vicinity $V_{1,1}$ of $S_{1,1}$, we open the tubes with a radius smaller or equal to $r_{min,1}$ in $V_{1,1}$, to obtain $S_{1,2}$ (step 1). As for drainage, we continue the process until the set of opened tubes stabilizes (last step). The water saturation associated to $r_{min,1}$ is also computed using Eq. (3). Then, considering the vicinity $V_{2,0} = V_{1,N1}$, we opened in it the smallest tubes, of diameter $r_{min,2}$ (which is larger than $r_{min,1}$), and repeat the procedure from step 1 to the last step, and so on, until the network is totally opened and thus fully saturated. Note that we do not consider air trapping here: the saturation process is similar to what we could obtain by putting the end face $y = 1$ under high fluid pressure or by first saturating the medium with gas that is easy to dissolve in the water, as it is often done in soil physics. Examples of saturation states are shown in Figure 5. Again, there is a critical saturation for which the opposite end-faces connect: around 8.38 % in the longitudinal direction, and 13% for the transversal direction.

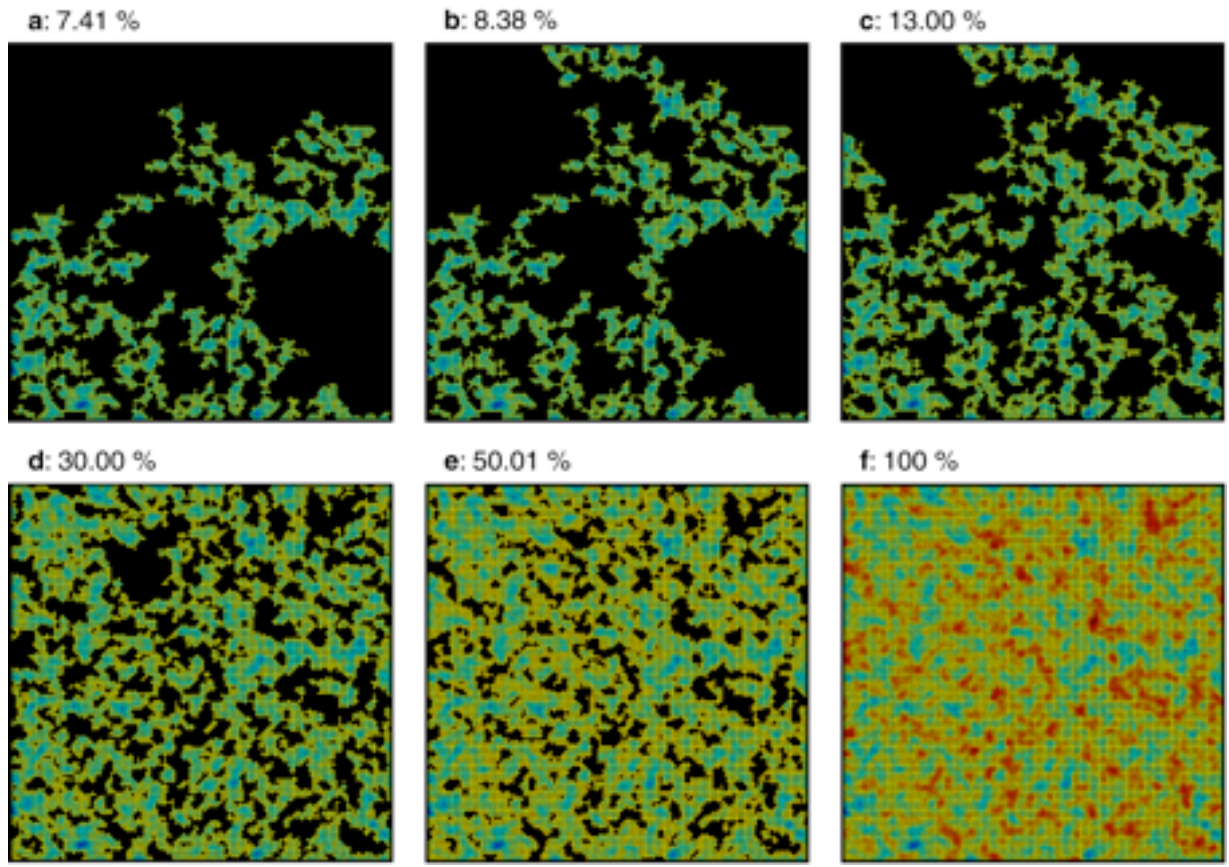

**Figure 5.** Example of saturation states during imbibition. The smallest radii are saturated first (i.e., all the radii where set to 0 at the initial state, and they are progressively set to their true value), starting from the bottom of the network (i.e., nodes $y = 1$, see Figure 1). The colour scale is the same as in Figure 2a.

**2.3 Relative permeability, resistivity index and characteristic lengths**

As pore network modelling is a well-known method (see for instance Kirkpatrick 1973; Koplik 1981; David *et al.* 1990; David 1993; or Bernabé 1995), we just give in the appendix a summary of the approach. Using it, we compute: (i) the relative permeability $k_r$ of the medium (the permeability at partial saturation being given by $k_r k$, with $k$ the permeability at full saturation), (ii) the resistivity index *RI*, defined as the ratio of the electrical resistivity of the medium at partial saturation divided by the resistivity at full saturation (Archie 1942) – note that we have neglected the surface conductivity in computations, i.e., only the conductivity of the fluid is taken into account, and (iii) The characteristic lengths $\Lambda_h$ (hydraulic) and $\Lambda_e$ (electrical) (in m) (Johnson *et al.* 1986, Schwartz *et al.* 1989; Bernabé &

Revil 1995), which can be seen as effective pore radii, and which are widely used parameters in petrophysics to link electrical conductivity and permeability (e.g. Schwartz *et al.* 1989, Bernabé & Revil 1995). In an electrical conductivity experiment, the length scale $\Lambda_e$ denotes an effective pore radius weighted by the norm of the nornalized electrical field in absence of surface conductivity (i.e., weighted by the local electrical field nornalized by the applied electrical field in absence of surface conductivity, e.g., Avellaneda & Torquato, 1991, and references therein).

**2.4 Computation of the SIP spectra**

To compute the SIP spectra of the networks at different saturation states, we used exactly the same approach as Maineult *et al.* (2017), and we modify the numerical scheme to account for tubes having radius equal to 0 (i.e. local impedance equal to 0). The local impedance of each tube of radius $r$ non equal to 0 is given by a Warburg resistivity model written here in its complex conductivity form (Revil *et al.* 2014b; Niu & Revil, 2016):

$$\sigma_l^* = \sigma_{0,l}\left(1 + \frac{m_l}{1-m_l}\left(1 - \frac{1}{1+(i\omega\tau_l)^{1/2}}\right)\right) \tag{4}$$

with

$$\tau_l = \frac{r^2}{2D_l} \tag{5}$$

Note that this model is equivalent to a Warburg conductivity model using the relationship found in Tarasov & Titov (2013) between such fonctions. The rational for a Warburg mode has been explored in details in Revil *et al.* (2014b) and Niu & Revil (2016) and will not be repeated here. The conductivity $\sigma_{0,l}$ denotes the DC conductivity (in S m$^{-1}$), $m_l$ is the chargeability (dimensionless, comprised between 0 and 1), equal to $(\sigma_{\infty,l} - \sigma_{0,l}) / \sigma_{\infty,l}$ where $\sigma_{\infty,l}$ is the limit of $\sigma_l^*$ for infinite frequency, $\tau_l$ is the time constant (in s), and $D_l$ is the diffusion coefficient of the counterions (Revil *et al.* 2012; Niu & Revil 2016). When a tube has a radius equal to 0 (desaturated state), we obtain $\sigma_l^* = 0$ by imposing $\sigma_{0,l} = 0$. As done by Maineult et al. (2017), we simplify the problem by imposing $\sigma_{0,l} = 0.01$ S m$^{-1}$, $m_l = 0.1$ and $\log_{10}(D_l) = -11$ for all tubes with non-zero radius.

Once all the local impedances are attributed, we compute the complex conductivity spectra. Then we fit them with a macroscopic Pelton model in conductivity, which expresses the bulk (macroscopic) complex conductivity as:

$$\sigma_{bulk}^* = \sigma_0 \left( 1 + \frac{m}{1-m} \left( 1 - \frac{1}{1+(i\omega\tau)^c} \right) \right) \quad (6)$$

where $\sigma_0$, $m$, $\tau$ and $c$ are the macroscopic Cole-Cole parameters, the Cole-Cole exponent $c$ being comprised between 0 and 1.

## 3 RESULTS

### 3.1 Relative permeability, resistivity index, and characteristic lengths

The matric potential $h$ (computed using Eq. 2 for all radii) as a function of the saturation $S_w$ exhibits a hysteretic behaviour (Figure 6) as observed in real media (e.g. Mualem 1973, Pham *et al.* 2005). Hysteresis is also observed for the relative permeability (Figure 7a), even though the hysteretic behaviour is less marked when the relative permeability is measured in the transversal direction. The relative permeability decreases with saturation, with the values computed during imbibition always larger than during drainage. Concerning the resistivity index (Figure 7b), it increases with decreasing saturation, with the values computed during drainage larger than during imbibition. Moreover the measurements in the transversal and longitudinal direction are relatively close one to each other for saturation below 0.3. The hysteretic behaviour of the resistivity index computed along the longitudinal direction is very similar to those of the resistivity measurements reported by Knight (1991) for sandstone samples. We also observe a hysteretic behaviour of the characteristic lengths (Figure 7c). Interestingly, the relationship between the characteristic lengths and the relative permeability seem to have two regimes (Figure 8): for relative permeabilities lower than 0.002, the characteristic lengths do not vary with $k_r$; for larger values, we can observe a power-law relationship $\Lambda = \Lambda(S_w = 1)k_r^{\beta}$ with an exponent $\beta$ around 0.154.

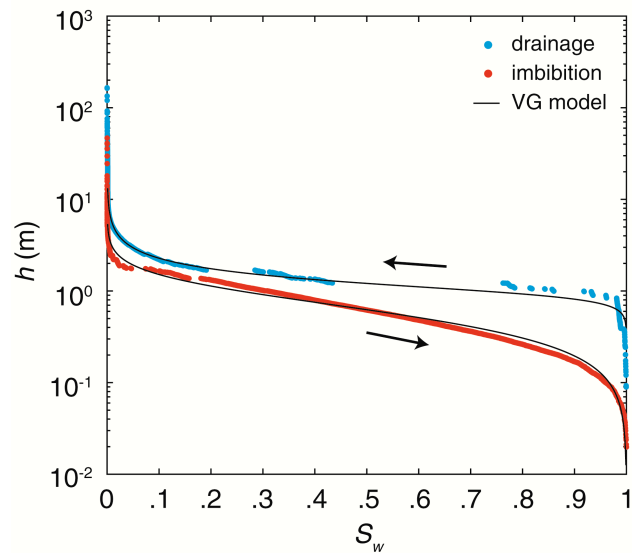

**Figure 6.** Representation of the matric potential $h$ in function of the saturation $S_w$. The hysteretic behaviour is well evidenced. Note that during drainage, the sudden connection of tube clusters triggers significant jumps in the saturation state. The black lines correspond to the Van Genuchten model obtained from independent fitting of the drainage and imbibition data (see Table 1).

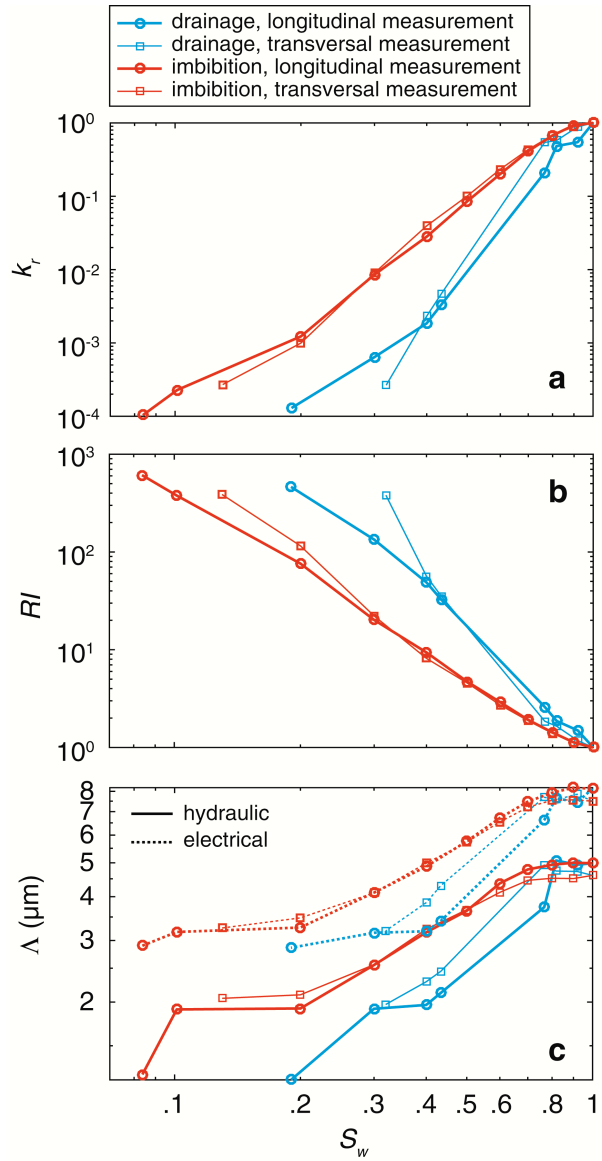

**Figure 7.** Evolution of the relative permeability $k_r$ (**a**), of the resistivity index $RI$ (**b**) and of the hydraulic and electrical characteristic lenghts $\Lambda_h$ and $\Lambda_e$ (**c**) as a function of the saturation of the network. All quantities behave hysteretically, except the relative permeability in the transversal direction (i.e., between nodes $x = 1$ and $x = N_x$, see Figure 1).

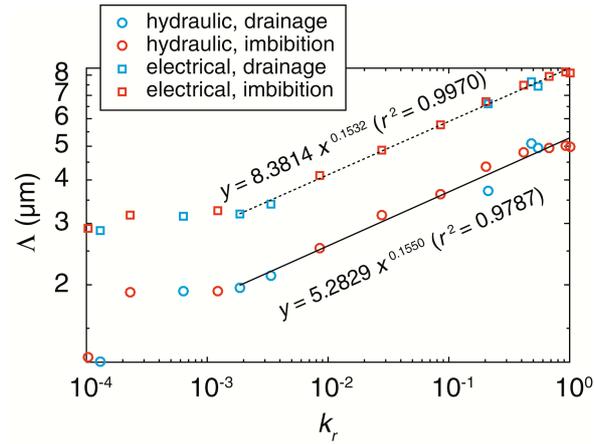

**Figure 8.** Evolution of the hydraulic and electrical characteristic lengths $\Lambda_h$ and $\Lambda_e$ in function of the relative permeability $k_r$.

### 3.2 SIP spectra

We observe a marked decrease of the amplitude of the complex resistivity during drainage, in both longitudinal and transversal directions (Figure 9a, c). With decreasing saturation, the phase spectrum tends to sharpen; its peak frequency (i.e., the frequency associated with the maximum value of the phase) moves towards higher frequencies, and its amplitude increases (Figure 9b, d). During imbibition (Figure 10), the spectra follow the opposite behaviour, i.e., an increase for the amplitude of the complex resistivity, a broadening of the phase curve, with a decrease of the peak frequency and a decrease of the phase amplitude.

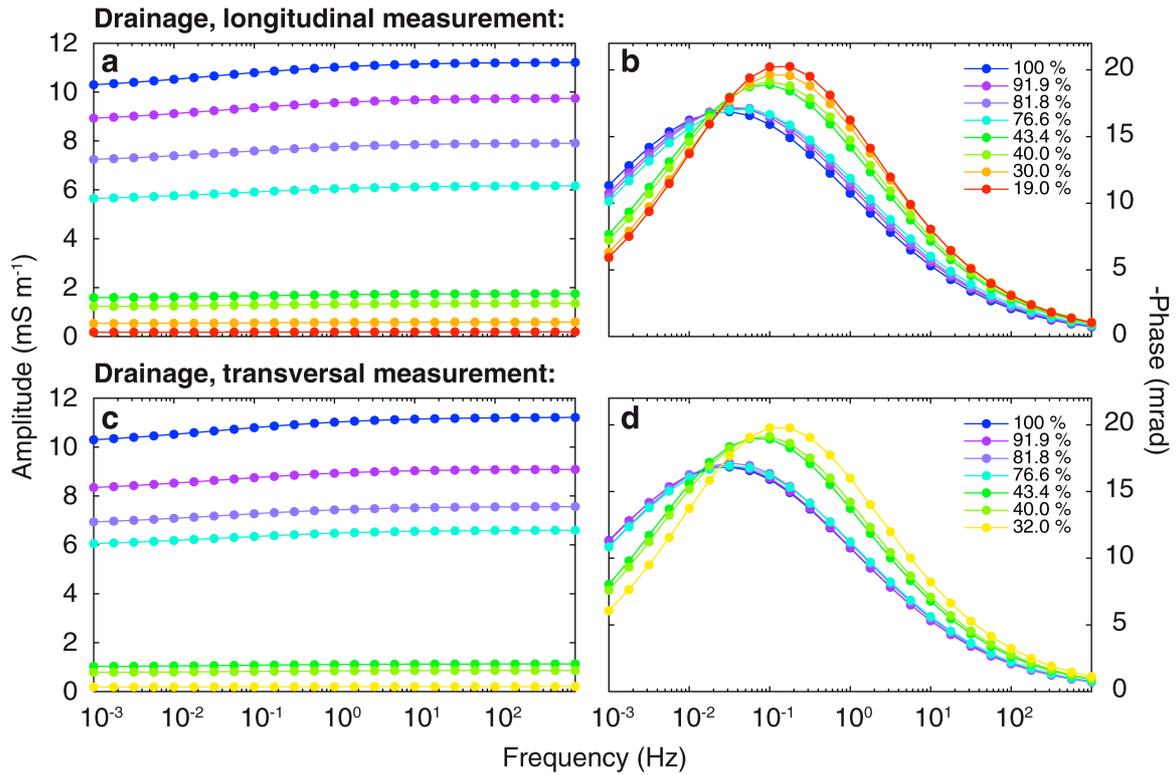

**Figure 9.** Complex conductivity spectra (amplitude and phase) at different states of saturation for the drainage, measured in the longitudinal direction (i.e., between nodes y = 1 and y = Ny, **a**,**b**) and in the transversal direction ((i.e., between nodes $x = 1$ and $x = N_x$, **c**, **d**). Note the decrease in amplitude spectra, and the increase of the frequency peak in the phase spectra with decreasing saturation.

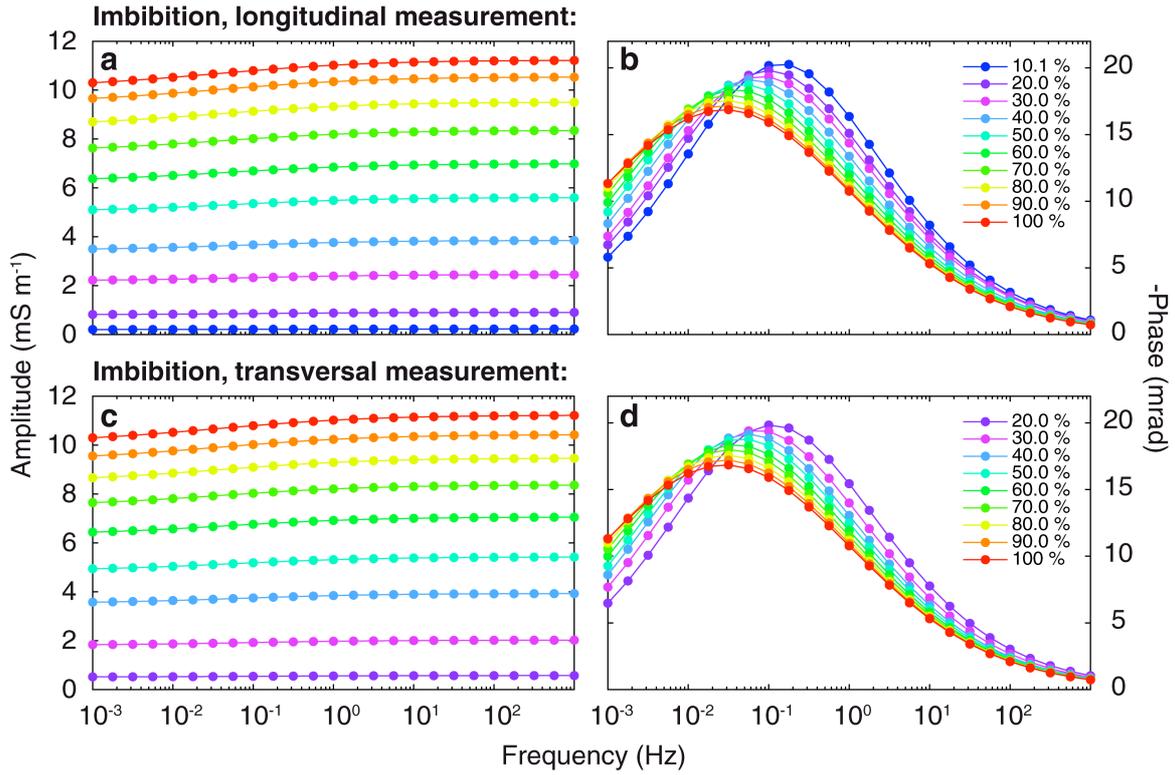

**Figure 10.** Complex conductivity spectra (amplitude and phase) at different states of saturation for the imbibition, measured in the longitudinal direction (i.e., between nodes y = 1 and y = Ny, **a**,**b**) and in the transversal direction ((i.e., between nodes $x = 1$ and $x = N_x$, **c**, **d**). Note the increase in amplitude spectra, and the decrease of the frequency peak in the phase spectra with increasing saturation.

All the SIP curves can be fitted by a Pelton model in conductivity (Eq. 6), with very good fit. It is thus possible to draw the evolution of the model parameters with saturation, i.e. the DC conductivity $\sigma_0$ (Figure 11a), the chargeability *m* (Figure 11b), the decimal logarithm of the time constant $\tau$ (Figure 11c) and the Cole-Cole exponent *c* (Figure 11d). All of them exhibit a hysteretic behaviour with respect to the saturation, in both longitudinal and transversal direction, even though it is less marked for the Cole-Cole exponent.

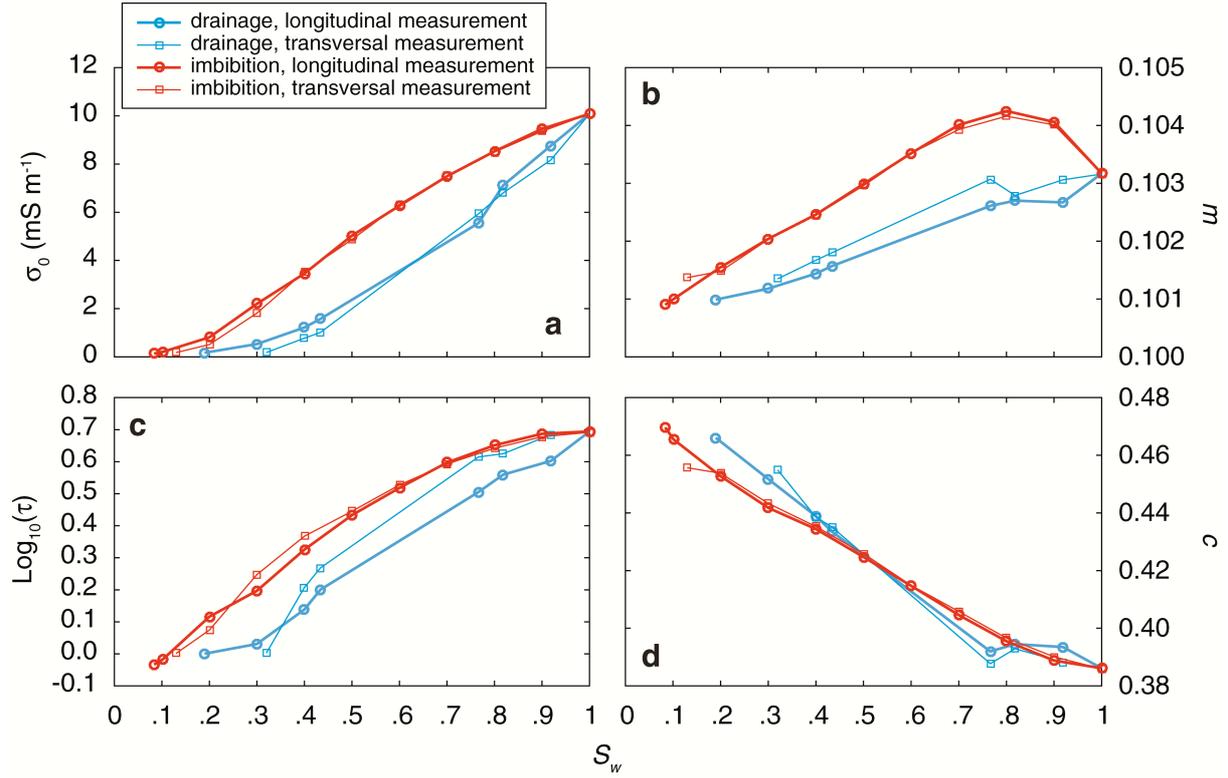

**Figure 11.** Evolution of the macroscopic Cole-Cole parameters : DC conductivity $\sigma_0$ (**a**), chargeability $m$ (**b**), time constant $\tau$ (**c**) and Cole-Cole exponent $c$ (**d**) in function of the saturation $S_w$, obtained by fitting the macroscopic complex resistivity spectra (Figures 9 and 10). All parameters exhibit a hysteretic behaviour, even though it is less clear for $c$.

## 4 DISCUSSION

### 4.1 Hydrodynamic

Concerning the evolution of the matric potential, we fit these two curves using van Genuchten's (1980) model (Figure 6):

$$h = -\frac{1}{\alpha}\left[\left(S_w^{-\frac{1}{m_{VG}}}\right) - 1\right]^{\frac{1}{n_{VG}}}. \tag{7}$$

Note that, in order to be able to fit the imbibition curve, it was necessary to decouple the two textural parameters as $m_{VG} \neq 1 - n_{VG}^{-1}$. In order to be consistent we optimized the three parameters $\alpha$ (in m$^{-1}$), $m_{VG}$, $n_{VG}$ for both the drainage and imbibition curves, the best fit parameters are presented in Table 1. As expected from the literature, at a given saturation, the matric potential obtained during the drainage is higher than the one during the imbibition (e.g., Mualem 1984, his Fig. 1). This results in a lower value of the optimized $\alpha$ parameter for the imbibition curve than for the drainage one (Table 1).

**Table 1.** Best fit values obtained for the Van Genuchten model displayed on Figure 6. The Root Mean Square Error is computed using:

$$RMSE = \sqrt{\frac{1}{N}\sum_{i=1}^{N}\left[\log_{10}\left(h_i^{predicted}\right) - \log_{10}\left(h_i^{measured}\right)\right]^2}$$

| Model parameter[a] | Drainage | Imbibition |
|---|---|---|
| $\alpha$ (in m) | 1.048 | 0.768 |
| $m_{VG}$ (−) | 0.373 | 2.769 |
| $n_{VG}$ (−) | 7.083 | 1.7141 |
| *RMSE** (in m) | 0.0709 | 0.0832 |

The resistivity index denotes the ratio between the resistivity at full saturation divided by the resistivity at a given saturation $S_w$. We model the evolution of the resistivity index using second Archie's law (1942) neglecting in the present study the effect of surface conductivity. This yields:

$$RI = \frac{1}{S_w^n} \tag{8}$$

We find that the saturation exponent $n$ is equal to 4 for drainage and to 2.5 for imbibition (Figure 12a). Although $n = 4$ is fairly high and surprising for a natural porous medium (generally expected around $2 \pm 0.5$ in water wet rocks, see Archie, 1942 or Glover 2015), it could be linked to electrical tortuosity in 2D media. Indeed, Jougnot *et al.* (2016) obtained the same value while monitoring the electrical conductivity of a 2D synthetic medium submitted to various drainage procedures.

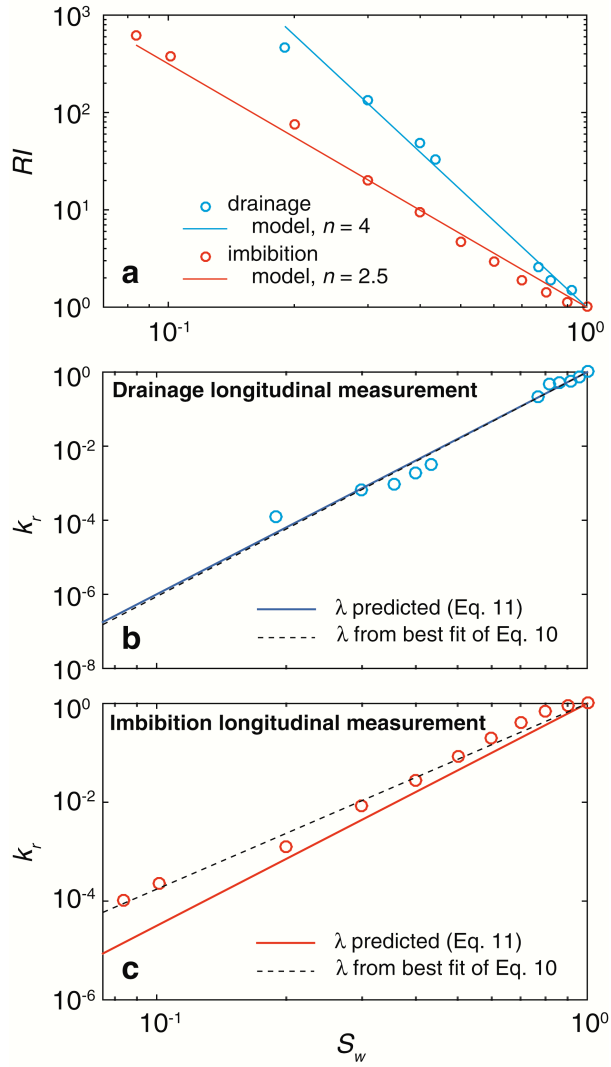

**Figure 12.** Modelling of the resistivity index $RI$ using $RI = S_w^{-n}$ for longitudinal measurements (**a**). The saturation exponent is equal to 4 for drainage and 2.5 for imbibition. Prediction of relative permeability $k_r$ with Brooks & Corey (1964) model (Eq. 10) using $\lambda$ value determined using Eq. 11 (continuous lines), and from best fit of Eq. 10 (dotted lines), for drainage (**b**) and imbibition (**c**).

We observe a hysteresis in the relative permeability (Figure 7a). This is apparently in contradiction with models, which predict that the $k_r(S_w)$ is non-hysteretic (e.g., Soldi *et al.* 2017), and with measurements (e.g., Topp & Miller 1966; Van Genuchten 1980; Mualem 1986). However, for saturations above 40 %, simulations show that the transversal section available for transport is more reduced during drainage than during imbibition (Figure 13, for saturation of 80 %), making the permeability also more reduced during drainage. In other words, since the geometry of the desaturated clusters are very different during drainage and imbibition, the flow paths are not similar, and so the resulting permeability. Note here that we were not able to fit the relative permeability curve versus saturation with classical models (in particular Brooks & Corey 1964, van Genuchten 1980). The only very good fit was obtained for the longitudinal drainage (for both Brooks & Corey (1964) and van Genuchten (1980) models, see below). For the three other curves, the best fit models were not able to reproduce them for the saturation range comprised between 0.6 and 0.95, as the relative permeabilities do not follow the expected power law function.

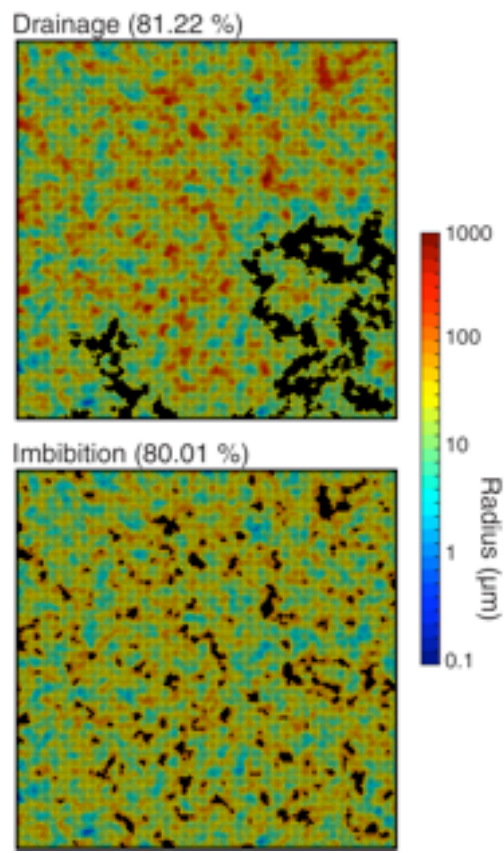

**Figure 13.** Network at about 80% of saturation, for drainage (top) and imbibition (bottom).

**4.2 SIP spectra**

The evolution of the SIP spectra with saturation (Figures 9 and 10) is very similar to those reported by Breede *et al.* (2011) and Breede (2012). Indeed, Breede *et al.* (2011) showed a decrease of the conductivity amplitude spectra with decreasing saturation for unconsolidated sand with negligible surface conductivity (around 5 µS cm$^{-1}$). Titov et al. (2004) also reported such a behaviour. Breede *et al.* (2011) and Breede (2012) also reported that the peak frequency slightly increased, the spectra are sharpened, and the maximum value of the phase increased with decreasing saturation. Also, on sandstone samples, Ulrich and Slater (2004) reported an increase of the phase at 1 Hz with decreasing saturation during drainage, and a decrease with increasing saturation during imbibition. Finally, Binley *et al.* (2005) observed an increase of the peak frequency with decreasing saturation on unsaturated consolidated sandstone samples. However, they did not observe a systematic increase in the maximum value of the phase, neither the sharpening of the spectra.

Concerning the Cole-Cole parameters, Titov *et al.* (2004) reported a decrease of the conductivity $\sigma_0$ with decreasing saturation for quartz-water-air mixtures, as we observe too (Figure 11c). For the chargeability, Breede *et al.* (2011) reported a decrease with increasing saturation for sand. Also for sand, Titov *et al.* (2004) observed an increase of the chargeability for saturation increasing between 0 and 0.1 and a decrease for saturation increasing between 0.1 and 1. Revil et al. (2012) reported a similar behaviour, with an increase of the chargeability for saturation below 0.1 and a decrease for saturation above 0.1. In fact, this seems to be the opposite of what we get (Figure 11b), in particular for drainage. For imbibition, we have such a behaviour, but with a change in slope at a saturation of 0.8, not 0.1. Concerning the time constant $\tau$, Breede *et al.* (2011), as well as Revil *et al.* (2012) observed an increase of the time constant $\tau$ with increasing saturation, as we do (Figure 11c).

**4.3 Link between SIP response and saturation and permeability**

Revil *et al.* (2014a) proposed new relationships to relate directly the electrical conductivity as a function of the saturation to relative permeability $k_r(S_w)$ through the use of the characteristic length Λ. They suggested that $\Lambda(S_w) = \Lambda(S_w = 1) S_w$, which yield to his model A:

$$k_r = S_w^{2+n}, \tag{9}$$

where $n$ is the saturation exponent as defined by Archie (1942) (Eq. 8). By comparing Eq. (9) to the Brooks and Corey (1964) model:

$$k_r(S_w) = S_w^{\frac{2}{\lambda}+3}, \tag{10}$$

where $\lambda$ is a textural parameter. Combining Eqs. (9) and (10) yields:

$$\lambda = \frac{2}{n-1}, \tag{11}$$

hence relating directly electrical ($n$) and hydraulic ($\lambda$) petrophysical parameters. By applying this to our data, where the saturation exponent $n$ is equal to 4 and 2.5 for the drainage and imbibition respectively, gives $\lambda$ equal to 0.667 and 1.333. Comparison between these predictions and best fit $\lambda$ values (from Eq. 10, $\lambda = 0.653$ and 2.633, respectively) are plotted on Fig 12b and c and reported in Table 2. Overall, these predictions of relative permeability curve from electrical conductivity measurements are fairly good but would need to be checked on other realizations of synthetic porous media.

**Table 2.** Predicted and best fit values to describe the relative permeability function using Eqs. (10) and (11)

|  | Saturation exponent, n | Predicted λ | Best fit λ |
|---|---|---|---|
| *Drainage* | 4 | 0.667 | 0.653 |
| *Imbibition* | 2.5 | 1.333 | 2.633 |

The relationship between the Cole-Cole DC conductivity $\sigma_0$ and the relative permeability appears to be a power law, with an early regime ($k_r < 0.01$) and a late one, for both drainage and imbibition (Figure 14a). A similar behaviour was reported by Breede (2012) on SIP measurements on pure sand and sand-clay mixtures. For the imbibition and longitudinal measurements, we obtain the relations $\sigma_0 = 52.056\, k_r^{0.6472}$ in the early regime, and $\sigma_0 = 9.9404\, k_r^{0.3039}$ in the late one (Figure 14b). Concerning the evolution of the time constant $\tau$ with respect to $k_r$, it is also a power law (Figure 14c), but it seems that there is only one regime, except for the imbibition with measurements in the longitudinal direction (Figure 14d). In this case, we find $\tau = 2.9498\, k_r^{0.1283}$ in the early regime, and $\tau = 4.9139\, k_r^{0.2392}$. This behaviour is very similar to what Breede (2012) measured (see her Figure 4.8): she obtained exponents in of the same order of magnitude (between 0.07 and 0.18). Finally, Binley et al. (2005) also reported a linear relationship between the time constant and the longitudinal hydraulic conductivity measured on sandstone samples, with exponent of 0.26 (their Figure 11).

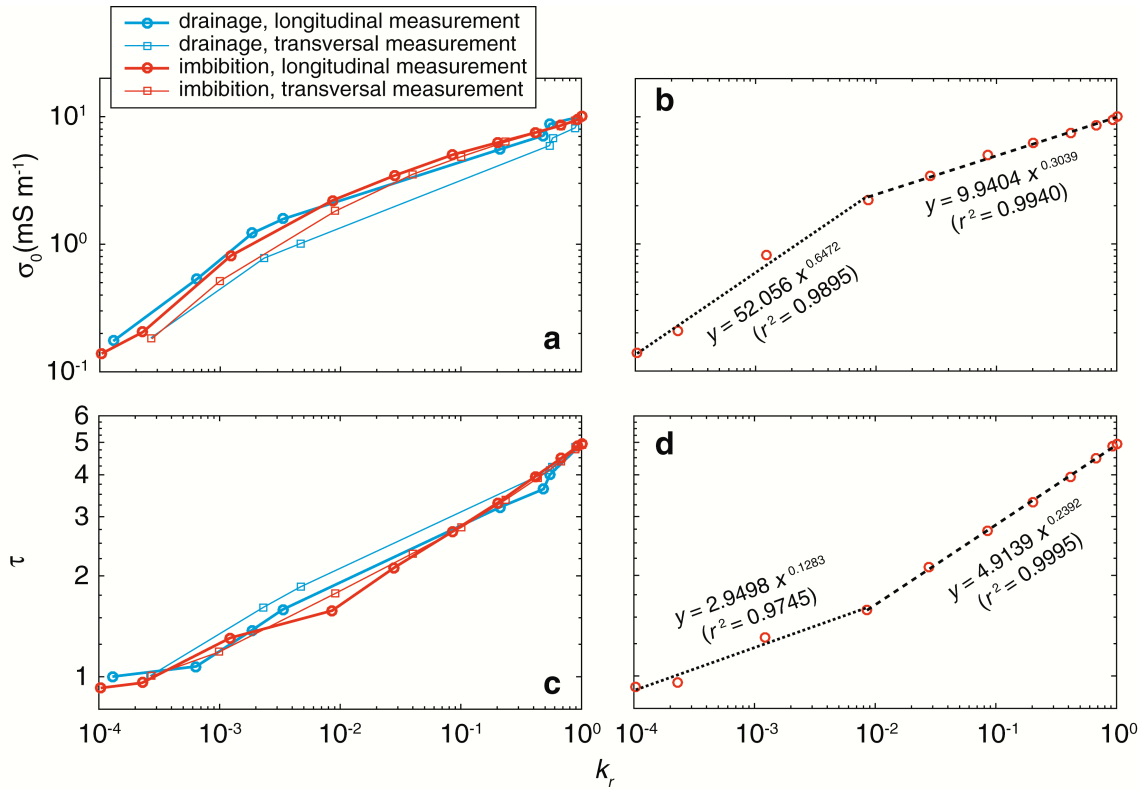

**Figure 14.** Evolution of the DC conductivity $\sigma_0$ (**a**,**b**) and time constant $\tau$ (**c**,**d**) in function of the relative permeability $k_r$. In **b** and **d**, only data for longitudinal measurements during imbibition are considered.

The Cole-Cole exponent $c$ seems to decrease linearly with increasing saturation (Figure 15), meaning that the value of $c$ may be a proxy for the saturation state, even though the coefficient of the slope is weak. One can notice that when decreasing saturation, the width of the radius distribution becomes narrower, i.e., less heterogeneous. Therefore the macroscopic value of $c$ tends towards the local value of $c_l$, a behaviour also reported by Maineult et al. (2017).

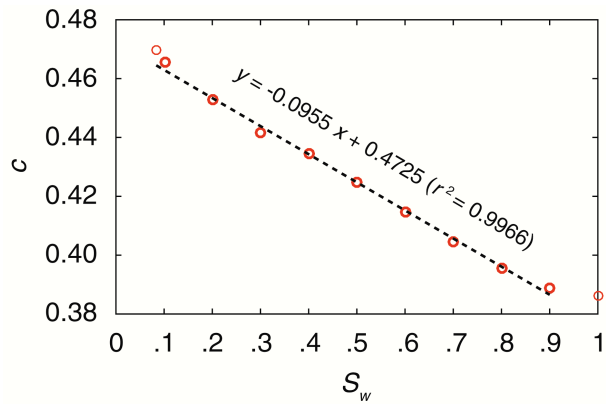

**Figure 15.** Evolution of the Cole-Cole exponent $c$ in function of the saturation $S_w$ during imbibition (longitudinal measurement). The relation is quasi linear, meaning that the saturation state could be deduced from $c$, even though the slope is rather small.

The characteristic lengths $\Lambda_h$ and $\Lambda_e$ decrease with increasing peak frequency $f_{peak}$ (the frequency for which the amplitude of the SIP phase spectrum is maximal), i.e., with decreasing saturation (Figure 16). The relationship between these quantities seems to be power-laws, with an exponent of about – 0.68. More interestingly, the peak frequency is related to the relative permeability also by a power law with an exponent of –0.185 (Figure 17). It means that SIP measurements could give access directly to the relative permeability, or, at least, that measured variations of the peak frequency due to saturation changes can provide information about the associated variations of the relative permeability.

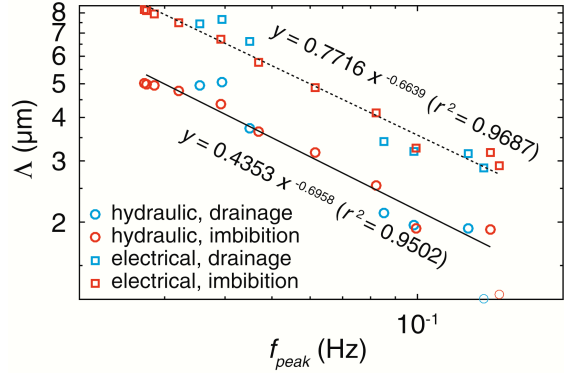

**Figure 16.** Evolution of the characteristic hydraulic and electrical lengths $\Lambda_h$ and $\Lambda_e$ in function of the peak frequency $f_{peak}$, for longitudinal measurements.

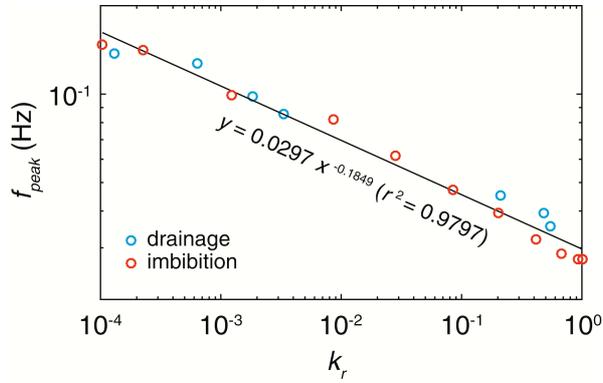

**Figure 17.** Evolution of the peak frequency $f_{peak}$ in function of the relative permeability $k_r$ (longitudinal measurements).

Obviously, only one simulation is not sufficient to draw general conclusions. That is the reason why we carried out similar calculations on four supplementary networks. In all cases, the behaviour of the relative permeability, the resistivity index and the characteristic lengths in function of the saturation was similar to the behaviour observed on the primary network (Figure 7). The complex conductivity spectra (Figures 9 and 10) were also similar, as well as the behaviour of the Cole-Cole parameters in function of the saturation (Figure 11). Concerning the relationships between the DC conductivity and the relative permeability, the previously described trend (Figures 14ab) is confirmed, with two power-law regimes (Figure 18a). Similarly, for the relationships between the time constant and the relative permeability, the previously described trend (Figures 14cd) seems to be also verified by the five simulations (Figure 18b). The linear trend between the Cole-Cole exponent and the saturation (Figure 15) is also observed (Figure 18c). Regarding the power-law relationship between the characteristic lengths and the relative permeability (Figure 8), it seems that in the general case it extends to the whole range of relative permeability (Figure 19a). Finally, the power-law regime between the characteristic lengths and the peak-frequency (Figure 16) can be generalized (Figure 19b), as well as the power-law regime between the peak frequency and the relative permeability (Figure 17 and Figure 19c).

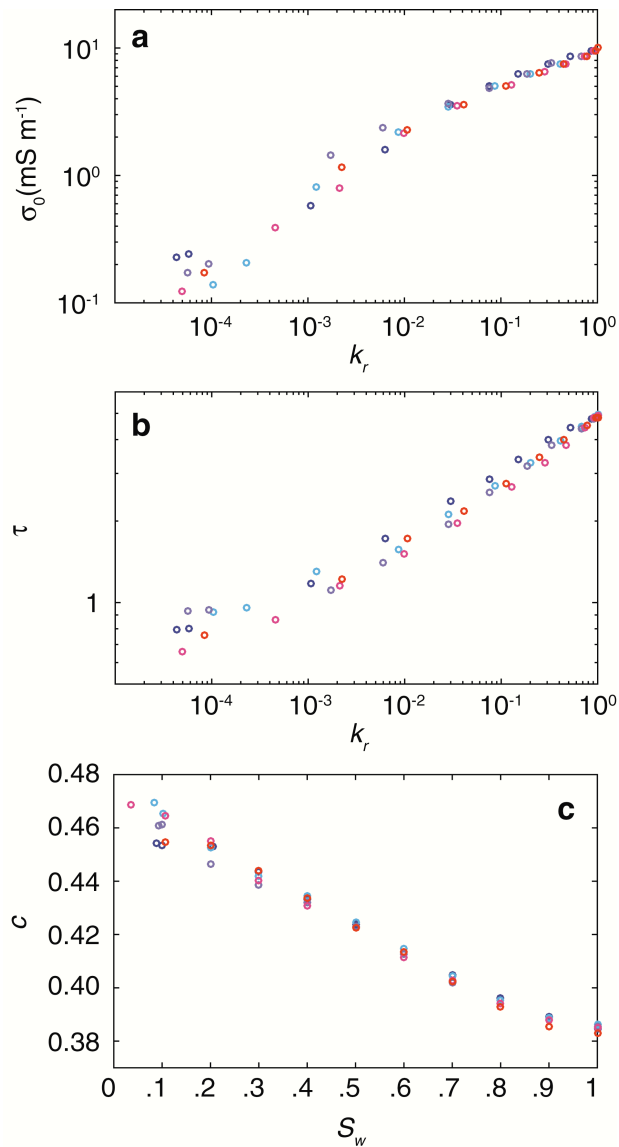

**Figure 18.** Results from 5 network simulations during imbibition. Evolution of the DC conductivity $\sigma_0$ (**a**) and time constant $\tau$ (**b**) in function of the relative permeability $k_r$ and evolution of the Cole-Cole exponent $c$ in function of the saturation $S_w$ (**c**).

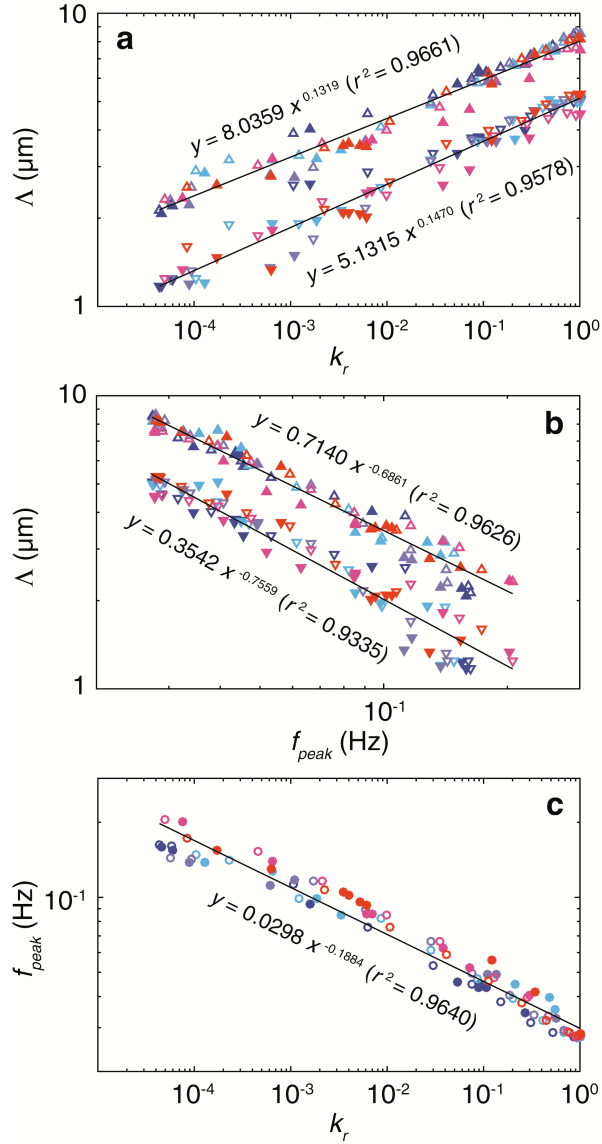

**Figure 19.** Results from 5 network simulations: evolution of the characteristic lengths $\Lambda_h$ and $\Lambda_e$ in function of the relative permeability $k_r$ (**a**) and of the peak frequency $f_{peak}$ (**b**), and evolution of the frequency peak $f_{peak}$ in function of the relative permeability $k_r$. ▼: $\Lambda_h$, ▲: $\Lambda_e$, plain symbols correspond to drainage and open symbols to imbibition.

Even though the representation of a porous medium by a correlated 2D square network is a very strong approximation, the simulations that we performed reproduced some behaviours previously reported by Breede (2012) on sands for instance, in particular the evolution of the DC conductivity and the time constant with the relative permeability. They also showed the potential interest of SIP measurements during drainage and imbibition to estimate some hydraulic parameters such as relative permeability from SIP parameters (in particular the peak frequency). Further works imply the extension of the method to 3D networks, for which the saturation and desaturation pathways should be somehow different, and also to test other types of mesh, since the transport properties of a network can depend on its topology (e.g., Bernabé *et al.* 2003). Finally, an experimental validation of the relationships that we deduced is highly advisable.

## 5 CONCLUSIONS

We are able to simulate saturation and desaturation processes in 2D square networks, and to calculate the associated resistivity index, relative permeability, characteristic lengths and SIP spectra (under the assumption that the parameters $\sigma_{0,l}$, $m_l$, and $D_l$ are the same for all tubes, as done by Maineult *et al*. (2017), and that the surface conductivity can be neglected). The behaviour of SIP spectra is globally in agreement with what was previously reported in the literature. In particular, for the phase spectra, there is an increase of the peak frequency with decreasing saturation. We also evidenced that the resistivity index, the relative permeability, the characteristic lengths and the macroscopic Cole-Cole parameters, except the Cole-Cole exponent *c*, present a hysteretic behaviour. We also show that the frequency peak is directly related to the characteristic lengths and to the relative permeability, meaning that permeability should be accessible from SIP phase spectra. As reported in the literature, we observe that the DC conductivity $\sigma_0$ and the time constant $\tau$ are related to the apparent permeability with a power law, meaning, here again, that quantitative information about the permeability can be deduced from Cole-Cole parameters. Finally, the Cole-Cole exponent is a quasi linear function of saturation. The computational exercise that we performed therefore evidences the interest of the SIP method for the study of the non-saturated medium and their evolution with water saturation. Further works implies the extension to 3D networks, the trapping of water in desaturated pores to account for the residual, irreducible saturation, as well as including the surface conductivity in the model.

# APPENDIX. RELATIVE PERMEABILITY, RESISTIVITY INDEX AND CHARACTERISTIC LENGTHS

We consider the network, saturated or not. The hydraulic flux $F_{p \to q}$ through a tube linking two nodes $p$ and $q$ writes:

$$F_{p \to q} = \frac{\pi r_{p \to q}^4}{8\eta} \frac{P_p - P_q}{l} = g^h_{p \to q} \left( P_p - P_q \right) \tag{A1}$$

where $r_{p \to q}$ is the radius of the tube and $l$ its length; $g^h$ is the hydraulic conductance, $\eta$ the dynamic viscosity, and $P$ the hydraulic pressure. To eliminate the length $l$, we introduce the modified hydraulic flux $\Phi^h_{p \to q}$:

$$\Phi^h_{p \to q} = F_{p \to q} l = \frac{\pi r_{p \to q}^4}{8\eta} \left( P_p - P_q \right) = \gamma^h_{p \to q} \left( P_p - P_q \right) \tag{A2}$$

Neglecting the surface conductivity, the electrical flux $J_{p \to q}$, taken writes:

$$J_{p \to q} = \sigma_w \pi r_{p \to q}^2 \frac{V_p - V_q}{l} = g^e_{p \to q} \left( V_p - V_q \right) \tag{A3}$$

where $\sigma_w$ is the electrical conductivity of the saturating fluid, $V$ the electrical potential and $g^e$ the electrical conductance. To eliminate $l$ and the fluid conductivity $\sigma_w$, we use the modified electrical flux $\Phi^e_{p \to q}$:

$$\Phi^e_{p \to q} = J_{p \to q} \frac{l}{\sigma_w} = \pi r_{p \to q}^2 \left( V_p - V_q \right) = \gamma^e_{p \to q} \left( V_p - V_q \right) \tag{A4}$$

At any node inside the square network (Figure 1), Kirchhoff's law (1845) writes:

$$Z_{x,y-1 \to x,y} + Z_{x-1,y \to x,y} + Z_{x+1,y \to x,y} + Z_{x,y+1 \to x,y} = 0 \tag{A5}$$

with $Z$ equal to $F$ or $J$ respectively. Using equation (A1) or (A3), this leads to:

$$\begin{aligned} &a_{x,y-1 \to x,y} X_{x,y-1} + a_{x-1,y \to x,y} X_{x-1,y} \\ &- \left( a_{x,y-1 \to x,y} + a_{x-1,y \to x,y} + a_{x+1,y \to x,y} + a_{x,y+1 \to x,y} \right) X_{x,y} \\ &+ a_{x+1,y \to x,y} X_{x+1,y} + a_{x,y+1 \to x,y} X_{x,y+1} = 0 \end{aligned} \tag{A6}$$

with $a = r^4$ and $X = P$ for the hydraulic case, and $a = r^2$ et $X = V$ for the electrical case. For the nodes on the border of the network, equation (A6) is easily modified to take into account the boundary conditions as described in Figure 1. A linear system is obtained; the $N_xN_y$ unknowns are the hydraulic pressure or electrical potential at the nodes of the network. Once this system is solved, the modified fluxes can be computed using equations (A2) and (A4). We used a QR decomposition to invert the matrix, which works even though the matrix is singular (for instance, when the four tubes connected to a given nodes are equal to zero).

The "apparent" permeability of the network is then computed using Darcy's law:

$$k_{app} = \frac{\eta Q L}{S|\Delta P|} = \frac{\eta}{l^2} \frac{N_y - 1}{N_x - 1} \frac{\Phi^h_{\Sigma out/in}}{|\Delta P|} \tag{A7}$$

where $L$ is the length of the network along the flow direction (i.e., $y$-direction), $S$ the transversal section, and the total out-flowing and in-flowing modified fluxes are given by:

$$\begin{cases} \Phi^h_{\Sigma out} = \sum_{x=1}^{N_x-1} \Phi^h_{x,N_y-1 \to x,N_y} \\ \Phi^h_{\Sigma in} = \sum_{x=1}^{N_x-1} \Phi^h_{x,1 \to x,2} \end{cases} \tag{A8}$$

For a fully saturated network ($S_w = 1$), $k_{app}$ is equal to the true permeability of the medium, denoted $k$. For unsaturated medium, it is convenient to introduce the relative permeability $k_r$, comprised and 0 and 1. It is defined as:

$$k_{app}(S_w) = k_r(S_w) k \tag{A9}$$

Therefore, introducing Eq. (A7) in Eq. (A9), the relative permeability is computed by:

$$k_r(S_w) = \frac{\Phi^h_{\Sigma out/in}(S_w)}{\Phi^h_{\Sigma out/in}(S_w = 1)} \tag{A10}$$

The "apparent" formation factor of the network is computed using:

$$\frac{1}{F_{app}} = \frac{\sigma_r}{\sigma_w} = \frac{1}{\sigma_w} \frac{JL}{S|\Delta V|} = \frac{1}{l^2} \frac{N_y - 1}{N_x - 1} \frac{\Phi^e_{\Sigma out/in}}{|\Delta V|} \tag{A11}$$

It is here convenient to introduce the resistivity index $RI$, defined by:

$$RI = \frac{\rho(S_w)}{\rho(S_w = 1)} \tag{A12}$$

where $\rho$ is the true resistivity of the medium. Replacing the resistivity by the conductivity, introducing the definition of the formation factor and using Eq. (A11), *RI* is computed by:

$$RI = \frac{\Phi^e_{\Sigma out/in}(S_w = 1)}{\Phi^e_{\Sigma out/in}(S_w)} \tag{A13}$$

Finally, we computed the two characteristic lengths (the first one, $\Lambda_h$, is based on the hydraulic potential, and the second one, $\Lambda_e$, on the electrical potential), as defined in Schwartz *et al.* (1989) using the formula given in Bernabé & Revil (1995):

$$\Lambda_h = \frac{\sum_{i=1}^{N_t} r_i^2 |\Delta P_i|^2}{\sum_{i=1}^{N_t} r_i |\Delta P_i|^2} \tag{A14}$$

and

$$\Lambda_e = \frac{\sum_{i=1}^{N_t} r_i^2 |\Delta V_i|^2}{\sum_{i=1}^{N_t} r_i |\Delta V_i|^2} \tag{A15}$$

where $\Delta P_i$ (resp. $\Delta V_i$) is the hydraulic pressure (resp. electrical potential) difference between the two end nodes of the $i^{th}$ tube.

**ACKNOWLEDGMENTS**

The authors sincerely thank the reviewers and the Associate Editor for their helpful comments.